\documentclass[epj]{svjour}
\usepackage{graphics}
\usepackage[T1]{fontenc}
\usepackage[latin9]{inputenc}
\usepackage{amsmath}
\usepackage{amssymb}
\usepackage{graphicx}
\begin{document}
\title{Growth of a tree with allocations rules: Part 1 Kinematics}
%\subtitle{Do you have a subtitle?\\ If so, write it here}
\author{Olivier Bui\inst{1} \and Xavier Leoncini\inst{1}% etc
% \thanks is optional - remove next line if not needed
\thanks{\emph{Present address:} Insert the address here if needed}%
}                     % Do not remove
%
          % Insert a name or remove this line
%
\institute{Aix Marseille Univ, Université de Toulon, CNRS, CPT, Marseille, France}
\date{Received: date / Revised version: date}
% The correct dates will be entered by Springer
%
\abstract{
A non-local model describing the growth of a tree-like transportation
network with given allocation rules is proposed. In this model we
focus on tree like networks, and the network transports the very resource
it needs to build itself. Some general results are given on the viability
tree-like networks that produce an amount of resource based on its
amount of leaves while having a maintenance cost for each node. Some
analytical studies and numerical surveys of the model in ``simple''
situations are made. The different outcomes are discussed and possible
extensions of the model are then discussed.
 % end of PACS codes
} %end of abstract
\maketitle

\section{Introduction}

Systems of transportation frequently appear in physics, engineering,
biology: transportation of water, electricity and gas in cities, street
network, river basins\cite{rodrig} or vascular systems\cite{Bohn2002,PhysRevE.73.061907}
etc. A typical way to formalize systems of transportation is to transform
it in a problem of flow in a network between sources and sinks\cite{ford62flows,aaarticle,ezvereefkpor}
with discussions about finding the best path from sources to sinks.
Such approach can lead to application in economic context\cite{Fulkerson59,Fulkerson1977,Hougaard2009}.
More physical approaches involving transport of fluids or electricity
have been considered too and, in these cases, use of the formalism
of electrical circuit (Kirchoff's law, current, potential etc.) is
not unusual\cite{PhysRevLett.98.088702,PhysRevLett.104.048703},
the more geometrical aspects would also be involved\cite{PhysRevE.73.016116,PhysRevLett.98.088701}.
This approach is often focused one minimizing some dissipative energy
which gives it some application in engineering context but also in
the description of nature as natural networks may arise from minimization
of some dissipation\cite{Rodriguez-Iturbe92,Banavar2001}.

We can notice that the focus is typically on optimization. The substance
being transported is ether not a resource to be used or consumed or
its use is of little importance to the problem the papers are studying.
A possible problem with such approach is that real life networks may
not just form from minimization/maximization problems but from some
interaction between to flow and the network: vascular system of plants
is a network transporting sugars, but these sugars are also used to
build the network itself. Perennial plants (including trees) do not
rely on a centralized organs such as a heart or a brain to grow or
distribute the resource throughout the plants and yet how they branch
and their final shape is not totally planned at birth either but is
also strongly caused by adaptation to external stimuli\cite{Chehab2008}.
Given this, we may believe trees are an example of self-organization
arising from very simple local rules and interactions\cite{SACHS2004197}.
The vascular system of trees is composed of a part called xylem and
another one called the phloem. The xylem is mainly responsible for
water transportation. This flow of water is unidirectional: it goes
from the roots to the leaves then most of it gets evaporated at the
level of leaves. These leaves then play a major role for the growth
and sustenance of the organism: they are the ones absorbing carbon
in the air and create sugars through photosynthesis. These sugars
are essential as they sustain the living cells of the tree as well
as provide its building materials. The part of the vascular system
responsible for transporting this vital resource is the phloem which
is the part we are interested with since it is the one that potentially
showcases this idea of interaction between the network and the resource
being transported. The biophysics of the phloem has been extensively
studied\cite{munch_stoffbewegungen_1930,DeSchepper2013,RevModPhys.88.035007}.

However, instead of studying the biomechanics of sugar transport like
the aforementioned papers, our approach is more abstract and in more
line with the work about flow, sources and sinks mentioned earlier
but we also propose to consider as well how the flow can be used to
built the network itself while in turn the flow depends on the network
and make a simple dynamical system out of it. In order to start on
relatively simple grounds the toy-model we built and present in this
paper will only use tree network (i.e. loopless). This has both the
advantage of keeping the network topology quite tractable, while potentially
shed some light on actual possible growth mechanisms occurring in
biological trees. The dynamical and growing aspects we added also
makes it similar to cellular automata which was a concept introduced
by John Von Neumann in \cite{von1996theory}, though a more famous
example of cellular automaton is the Game of Life invented by John
Conway\cite{gardner1970mathematical}.

In short, we are studying growing systems for which the growth and
self-organization is driven by short range interaction (exchange of
resource). The paper is organized as follows, in Sec.~\ref{sec:A-bio-inspired-model}
we discuss the ingredients of model inspired from trees, trying to
keep a minimal set of ingredients or variables. Then in the following
section \ref{sec:Example-of-implementation}, we propose a way and
implement the self-consistent dynamics of the growth of the tree.
In Sec.~\ref{sec:A-static-description} and Sec.~\ref{sec:Generalization-for-}
we perform some analytical calculations and predictions of the model
and then perform some more analysis with numerical simulations in
Sec.~\ref{sec:Simulation-of-symmetric} before concluding in Sec.~\ref{sec:Conclusion}.

\section{A bio-inspired model\label{sec:A-bio-inspired-model}}

\begin{figure}
\hfill{}\includegraphics[width=120bp]{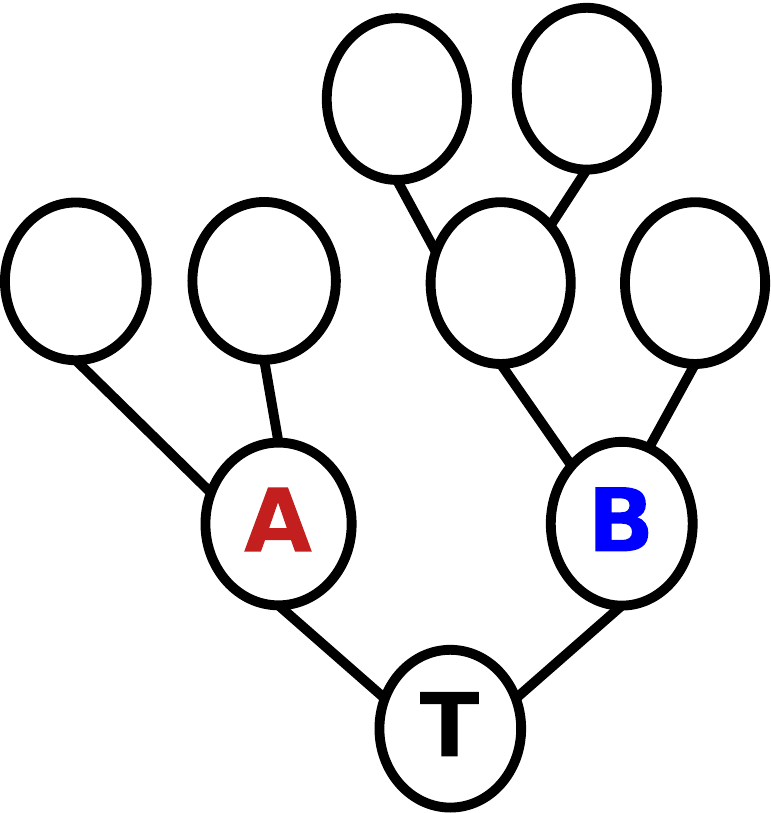}\hfill{}

\caption{\label{fig:Basic}Acyclic connected networks represent trees. To illustrate
the terminology we use in the paper: node T is the \textit{trunk},
it has two \textit{children} (node A and B which have two children).
The tree has 5 \textit{extremities} and is of \textit{height} 3. And
if the tree follows Leonardo's rule then the volume of T will be $5$
units of volume, A will be 2 units and B will be 3 units.}
\end{figure}

In this section we construct a ``growing'' system centered around
resource distribution and allocation inspired by biological trees.
Considering how complex trees are, we opted for what appeared to be
a simpler description of their growth and resource allocation. The
main constituent of a tree is carbon, it is absorbed by their leaves
and their source of energy is sunlight which also involve leaves.
Both the carbon and energy is distributed and allocated in the form
of sugars through the phloem. So for a simplistic description of trees
we may shave off the xylem, water transport and the roots, which leaves
us with a model featuring leaves as our sources, sugars as our resource,
branches representing the nodes of our growing transportation network,
and the volume of carbon each branch has fixated. Each branch will
need to regularly consume the resource (sugar) to increase in volume
(of carbon) and to keep its cells functioning. Let us introduce some
terminology for our model:
\begin{itemize}
\item \textit{Trees} are modeled as acyclic (i.e. loopless) connected networks
like in figure \ref{fig:Basic}, the nodes are called \textit{branches},
one of them is designated as the \textit{trunk} which is the only
node existing at the start of the simulation.
\item Each branch has a \textit{height} which is an integer indicating its
distance to the \textit{trunk} (the trunk is of height 0, the branches
directly linked to it are of height 1 etc). A branch with greater
height than another is said to be \textit{higher.}
\item If two branches are linked by an edge then the highest one is said
to be the \textit{child} of the second. The second is the \textit{parent}
of the first. The terms \textit{ancestors/descendants} will be used
for parent/children, grandparent/grandchildren, etc. And we say a
branch A \textit{descends} from another one B when the branch A is
a descendant of B.
\item Finally, we call \textit{extremities} branches without any children.
And the \textit{height of the tree} is defined as the height of the
highest extremity.
\end{itemize}
Now, we propose a few rules related to resource distribution and allocation:
\begin{itemize}
\item The extremities will be the \textit{sources} (the only branches possessing
leaves) and all the branches, including the extremities, need to consume
sugars as \textit{maintenance cost}.
\item Each branch can also \textit{use up its resource to increase its }volume,
the idea being large volume equates with more resilient branch, but
in exchange the \textit{maintenance cost is higher for a larger branch}.
\item Each branch can \textit{transfer resource to any of its children or
to its parent} and it can also store it in its \textit{``personal''
reserve}.
\end{itemize}
And the rules driving changes in the network are the following:
\begin{itemize}
\item If a branch \textit{dies} (could not pay its maintenance cost) then
all its descendants will die (the branch is cut off the tree). Since
the sources are at the tips of the tree, it creates an interdependence
between the ones producing resource and the rest of the tree.
\item About growth: extremities can \textit{use up resource to create children}
and in exchange cease to be sources while the newborns
will be the new extremities. It is through this process that the tree
can increase its number of sources since each time a branch creates
more than one child, the number of sources increases.
\end{itemize}
Time is discrete: the branches act in successive rounds. We will provide
later an example of how this ``game'' may progress.

The last characteristic we need to detail is the volume of branches.
Each branch possess their own quantity of volume (i.e. they have a
certain size). The bigger a branch is the more costly its maintenance
is, and to become bigger the branch also needs to spend resources.
In our model, we decide that the minimum volume (in some arbitrary
unit of volume) a branch needs to have is equal to the number of its
descendants that are extremities: if a branch has one child that is
an extremity, two grandchildren that are extremities and one great
grandchild that is an extremity, then the branch at least needs a
volume equal to 4 (in unit of volume). This rule will be called Leonardo's
rule because it is inspired from the real Leonardo's rule which has
been studied in plants\cite{leonardo_dav,SHINOZAKI1964KJ00001775191,MCMAHON1976443,PhysRevLett.107.258101}.

The exact formulation of the real Leonardo's rule is a rule of conservation
of cross section areas: when looking at a branch that branches out
into several others ones, the cross section of the branch before the
branching node is equal to the sum of the cross sections of the branches
after the node. Our Leonardo's rule would be synonymous with the real
one if consider the branches in our model as cylinder with all the
same length and the extremities of our tree all have a volume of 1
unit.

\section{Example of implementation\label{sec:Example-of-implementation}}

\begin{figure}
\begin{centering}
\includegraphics[width=4cm]{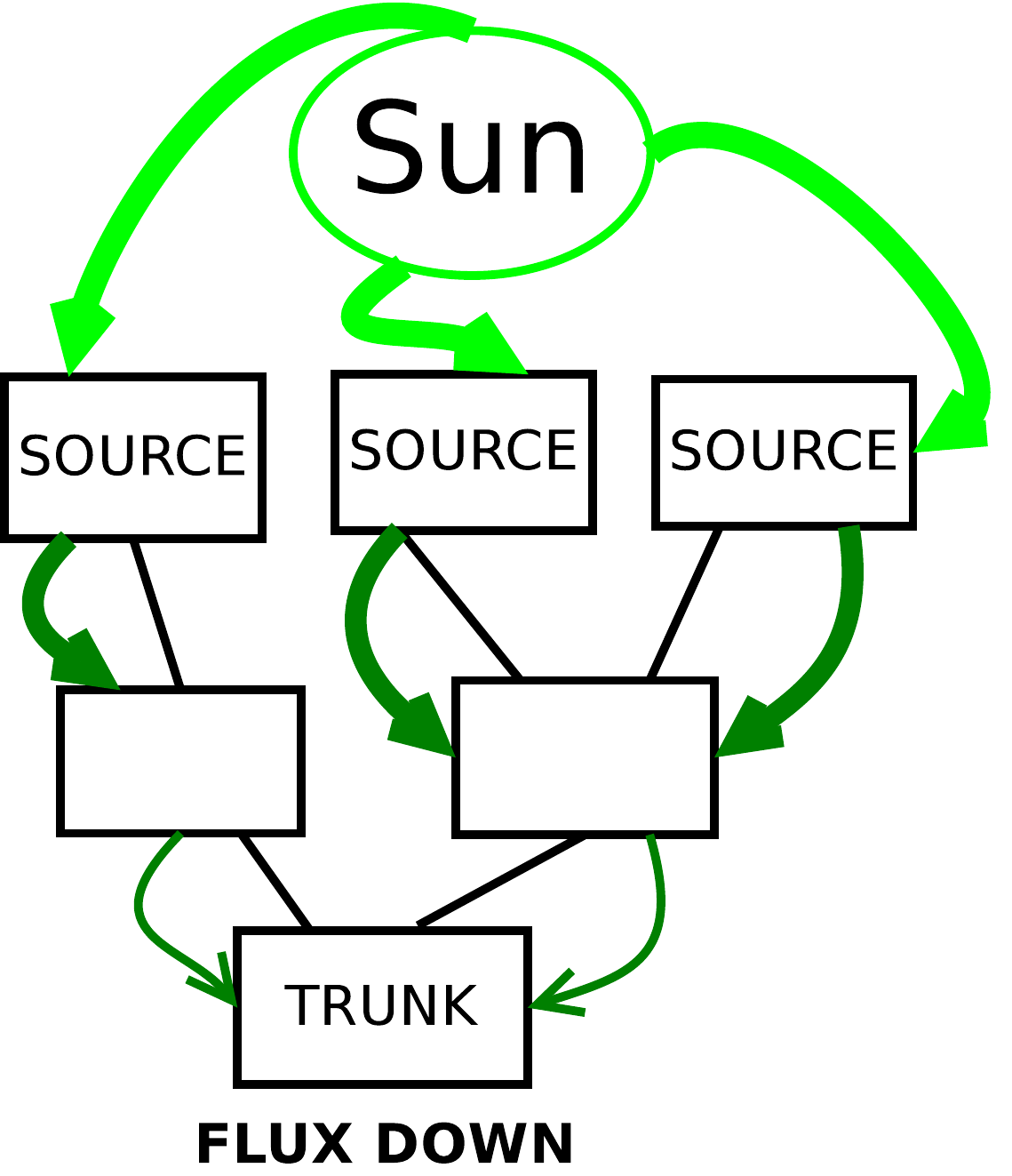}\includegraphics[width=4cm]{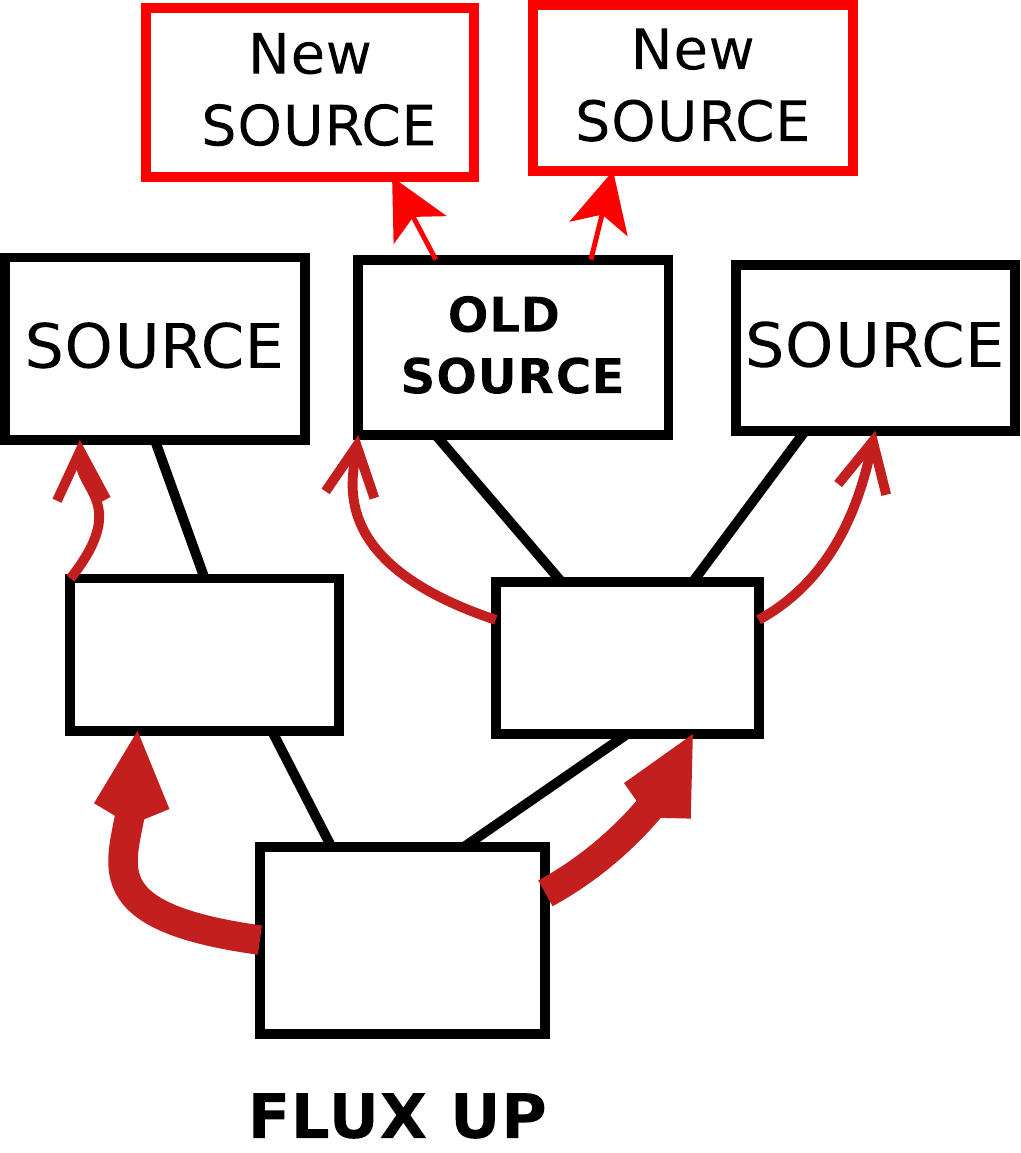}
\par\end{centering}
\caption{\label{fig:generation}The dynamics follows this sequence of events:
the sources get an amount of resource. Then they can choose to store
some amount for themselves and give the rest to their parent. Then
the parents do the same to the grandparents etc, until the flux reaches
the trunk. After this, the trunk pay its maintenance cost and use
some of the resource to grow in volume, then divides the remaining
to its children. The process is repeated until we reach the extremities
with no child, these last one can use up the remaining resource to
birth new branches. During the phase where the flux goes upward called
``flux up'', branches that cannot pay their maintenance or grow
to the correct volume may die.}
\end{figure}

We have established the basic rules describing our model. However
such description does not specify the model we implement. We still
have to describe how the ``game'' progresses in time or, in other
words, the dynamics of the system.

Example of a possible dynamics: we start with an amount of resource
created at each source then during each ``turn'' every branches
can store, consume for volume growth or transfer any amount of resource
to their neighbors. And then after $x$ number of ``turns'' each
branch has to pay maintenance from their stored resource or die and
we repeat process. Creating new branches may involve consuming resource
for $y$ numbers of unit of time etc.

Such dynamics would be fairly difficult to study, so we opted for
a simpler dynamics. Unlike the example described above, in our chosen
version, the branches do not act simultaneously but in succession.
We will call this succession of events and actions a \textit{generation}.
When a generation ends, a new one begins, so generations will be used
as our unit of time. Each generation is composed of two phases and
each branch acts once during each phase. The first phase describes
how the amount of resource generated at the extremities flows ``downward'':
from the extremal branches to eventually the trunk. And the second
phase describes how the resource reaching the trunk bounces back ``upward'':
from the trunk to the extremities.

\begin{figure}
\hfill{}\includegraphics[width=100bp]{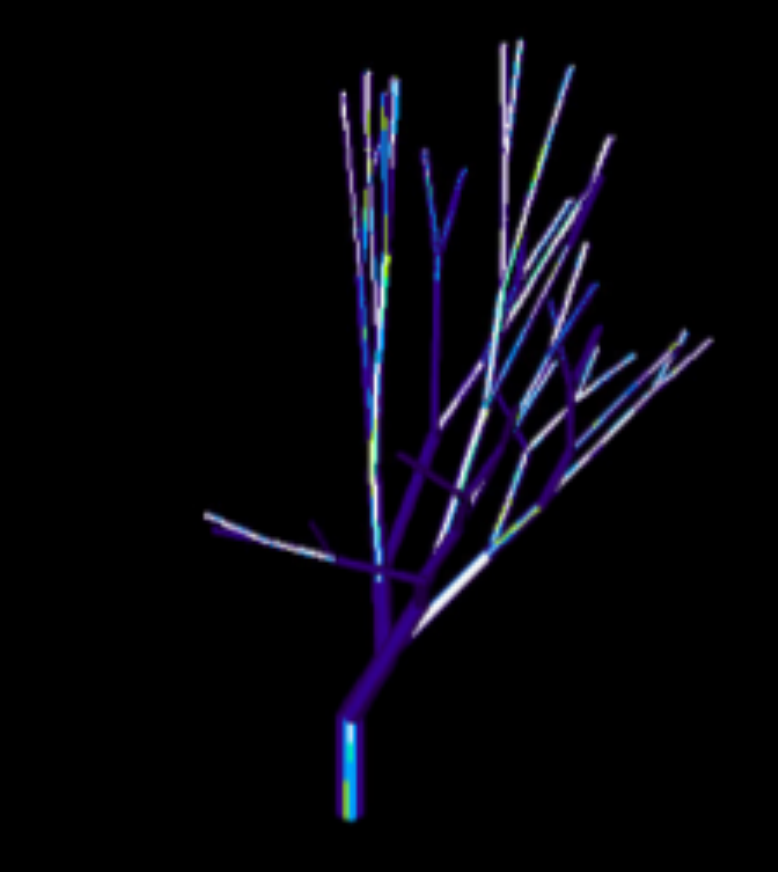}\hfill{}

\caption{\label{fig:sim} A 3D rendering of a tree created from the model.
The angle of the branches were chosen to make the drawing visually
appealing but in our model trees exist as pure graphs free of spacial
constraint.}
\end{figure}

Let us specify this first phase which we call ``flux down'' (Fig.~\ref{fig:generation}):
the first branches to \textit{act} are the extremities, then the next
to \textit{act} are the branches for which all the children have already
\textit{acted}, and we repeat until every branches have \textit{acted}.
Each \textit{action }is the following sequence of events:
\begin{itemize}
\item If the branch is an extremity then it receives an amount $p_{0}$
of resource. Otherwise, it receives an amount from its children and
will remember which amount was received from which child.
\item It also receives from each child information about their energetic
needs which is their maintenance cost as well as the resource they
need to create the additional volume of wood necessary to support
the structure above them. This information about energetic needs determines
how the resource will be distributed during the second phase when
the parent will have to choose the amount to distribute to each child.
\item The branch can keep some of the resource in its reserve then transfer
the rest to its parent. Then, it is this parent that will perform
this same sequence of events.
\end{itemize}
In the second phase which we call ``flux up'', the branches also
act in succession but the order is reversed (Fig.~\ref{fig:generation})
and the actions are now:
\begin{itemize}
\item If the branch is the trunk then it will start with all the resource
that was given to it during flux down, then the flux will bounce back.
If it is not the trunk it will have the resource it kept in its reserve
but has also received an amount from its parent.
\item With this amount of resource, it must pay its maintenance cost which
depends of its volume and dies if it can not.
\item Then, it will try to use up resource to grow to its ``optimal''
volume which is equal to the number of extremities that are its descendants
(the volume of a newborn branch being set to one unit of volume, it
means if the volume of each branch is ``optimal'', in the sense
we have defined above, then the tree will perfectly respect Leonardo's
rule). If the branch volume is not larger or equal to its optimal
volume, it is cut off.
\item The branch shares all its remaining resource to its children. The
share given to each one depends on what happened during the flux down
phase: the energetic needs each one told as well as the amount of
resource each gave.
\item If the branch has no child (it is a source) then it uses
the resource to create children. There is a maximum number of children
it can create.
\end{itemize}
In this phase, the proportion a parent shares/distributes its resource
to one child or another is decided by a formula of our choosing. However
the parent is supposed to remember the amount of resource each child
gave during flux down as well as their energetic need. So a suitable
formula would be one that use these two values, doing so we obtain
the tree represented in Fig.~\ref{fig:sim}.

In the end the dynamics of the tree only depends on a few parameters: 
\begin{enumerate}
\item The quantity produced by one extremity $p_{0}$.
\item The cost of creation of a new branch $\mathcal{C}$.
\item The maintenance cost of a branch of volume $V$: $m_{0}\times V{}^{\alpha}$.
\item Cost of creating more volume of wood (increasing the branch's diameter).
The  cost is linear $C_{v}\times(\textrm{Vol created})$.
\item The maximal number of children a branch can create $N_{max}$.
\end{enumerate}

Using the simple dynamics we described, we can obtain the tree that
was represented drawn in three dimensions in Fig.~\ref{fig:sim}.
All the branches were identical: during flux down, they all use the
incoming resource using the same algorithm (they grow in volume in
order to get the volume dictated by Leonardo's rule) during flux up,
they all use the same algorithm to decide how they share their resource
among the children (the algorithm take as entries the energetic need
of a child and its previous contribution during flux down) but with
a very small proportion of the resource being sometimes distributed
randomly. As for the branches with no child, they would use the remaining
resource to create as many children as possible.

On the other hand, if, during flux up, we do not implement that slight
random distribution then the branches will always split equally their
resource among their children because every branch have the same strategy
so, during flux up redistribution of resource to the children, the
parent will have no reason to discriminate among the children. This
results in ``symmetric'' trees. Either way, the growth and trees
we obtain depend on the parameters ($p_{0}$, $\mathcal{C}$, $m_{0}$,
$C_{v}$, $\alpha$ and $N_{max}$). Therefore, we may want to have
a theoretical view on the system. The simplest way to look at the
system is by having a static approach: for instance, for a tree with
a given topology we can calculate the total resource produced and
its total maintenance cost assuming the volume of the branches follows
Leonardo's rule.

\section{A static description of the problem for the case $\alpha=1$\label{sec:A-static-description}}

If we solely look at the total production and maintenance cost of
a tree for a given topology, it will allow us to make statements that
are independent of the dynamics we chose, as such, this is the approach
we will start with. For the calculations in this Section, the only
hypothesis that will be used are: every extremities produce the fixed
amount $p_{0}$ per unit of time, each branch pays the cost 
$m_{0}\times(\textrm{Vol branch})^{\alpha}$
per unit of time and the volume of the branches are such that they
respect Leonardo's rule, i.e no branches have been cut above it.

We may want to interpret $\alpha$. If we were look at our tree in
the context of botany: the branches in our model are supposed to be
cylinders of same length and the only morphological difference between
the branches is their radius, therefore the volume scales like its
cross-section surface. $\alpha=1$ means the maintenance cost is $m_{0}\times(\textrm{Vol branch})$:
this proportionality to the volume can be interpreted as the cells
being uniformly spread in the branch and consuming resource at the
same rate. In reality, for large trees and branches, living cells
are thinly located at the exterior of the branch while the interior
of the branch is mostly dead wood: this would correspond to $\alpha\simeq1/2$.
Outside the context of botany an analogy to be drawn between our cost
maintaining a structure (volume) and the results on dissipative cost
of transporting resource found in \cite{PhysRevLett.84.4745}. (The
comparison mainly holds thanks to the conservation of volume i.e.
Leonardo's rule.) Their paper found trees (loopless network) as the
``optimal'' topology when $0<\alpha<1$ but not in other cases.
Considering this fact, we will ignore $\alpha>1$. The case limit
$\alpha=1$, however, can still give us some insight for $0<\alpha<1$
and makes analytical calculations easier.

Since we assume all extremities (and only them) create an amount of
resource $p_{0}$, then the total production, noted $\mathcal{P}$,
is equal to $p_{0}\times E$ where $E$ is the total number of extremities
the tree has. Now, we recall that, for each branch, its height is
an integer defined as the distance between the branch and the trunk
within the graph representing the tree. For each extremity of the
tree, we can note its height, then we define $E_{h}$ as the number
of extremities that have a height equal to $h$, as a result $E=\sum E_{h}$.
Thus, by defining $H$ as the height of the highest extremity, we
can write: 
\begin{equation}
\mathcal{P}=\sum_{h=0}^{H}E_{h}\times p_{0}\label{eq:first}
\end{equation}

We call $\mathcal{M}$ the total maintenance cost. It has a simple
expression when $\alpha=1$ (the linear case): 
\begin{equation}
\mathcal{M}={\displaystyle \sum_{h=0}^{H}}(h+1)\times E_{h}\times m_{0}\label{eq:ee}
\end{equation}
To illustrate this formula we can look back at Fig.\ \ref{fig:Basic}:
if Leonardo's rule is respected the trunk has a volume of 5 because
there are 5 extremities that are its ``descendants''. Again with
Leonardo's rule, if we look at the children of the trunk (the branches
named A and B): one has a volume equal to 2 and the other to 3. In
other words, summing the volume of the children of A and B also yields
5 which is the number of extremities that either ``descends'' from
A or from B. Then if we look the branches at the height just above,
the sum is also 5 which is the number of extremities descending from
them \textit{plus} the number of extremities among them. This last
reasoning yields 2 for the branches of height 3.

This observation can be generalized into the following rule: for a
tree of height $H$ and given any integer $j\in[0,H]$, the sum of
the volume of all branches that have a distance $j$ from the trunk
(height equal to $j$) is equal to $A_{j}=\sum_{h=j}^{H}E_{h}$. So
the total of volume of the tree is $\sum_{j=0}^{H}A_{j}$ which can
be rewritten as $\sum_{h=0}^{H}(h+1)\times E_{h}$. A way to interpret
this last expression is by saying that adding an extremity of height
$h$ increases the total volume by $(h+1)$ units. Thus, if the maintenance
cost of a branch is proportional to its volume (i.e. $\alpha=1$)
then, by linearity, we get the equation\ (\ref{eq:ee}) by multiplying
the total volume of the tree with the proportionality factor $m_{0}$.
Hence for $\alpha=1$, by combining (\ref{eq:first}) and (\ref{eq:ee})
we obtain the global balance $\mathcal{B}$:

\begin{equation}
\mathcal{B}=\mathcal{P}-\mathcal{M}={\displaystyle \sum_{h=0}^{H}}E_{h}(p_{0}-(h+1)m_{0})\label{eq:alphaone}
\end{equation}

This expression for $\mathcal{B}$ is simple enough we can easily
deduce a few results. We recall $N_{max}$ is the maximum number of
children a branch may have. For a fixed $N_{max}<\infty$, and using
the expression (\ref{eq:alphaone}), we obtain the following results:
\begin{enumerate}
\item We can not find arbitrarily tall tree such that $\mathcal{B}\geq0$.
This in turn implies that no tree can grow infinitely in height as
they would reach a height above which $\mathcal{B}$ can only ever
be strictly negative.
\item For trees of height $H\leq H_{0}$, with $H_{0}$ being an integer
we will define later in Eq.~(\ref{eq:opti}), we know the form of
the tree maximizing $\mathcal{B}\geq0$: this maximal tree is the
tree for which every branches have $N_{max}$ children, except the
branches of height $H$. I.e. this is the tree of height $H$ that
has $N_{max}^{H}$, hence maximizing its number of extremities.
\end{enumerate}
In order to show this we start by defining the integer $h_{max}=\left[ \frac{p_{0}}{m_{0}}\right] -1$.
The brackets $\left[ x \right]$, indicate the integer part of $x$, and will be used as such hereafter.
This is the smallest integer for $h$ such that $p_{0}-(h+1)m_{0}$
is negative for any $h>h_{max}$. Therefore, in Eq.~(\ref{eq:alphaone}),
the terms of indexes higher than $h_{max}$ are negative. In other
words, any extremities added at a height strictly higher than $H_{max}$
will contribute negatively to our profits $\mathcal{B}$. Consequently,
since $N_{max}<\infty$, it is clear we can not find arbitrarily tall
trees such that $\mathcal{B}$ as we would reach a point where adding
branches will only contribute negatively.

\begin{figure}
\centering{}\includegraphics[width=5cm]{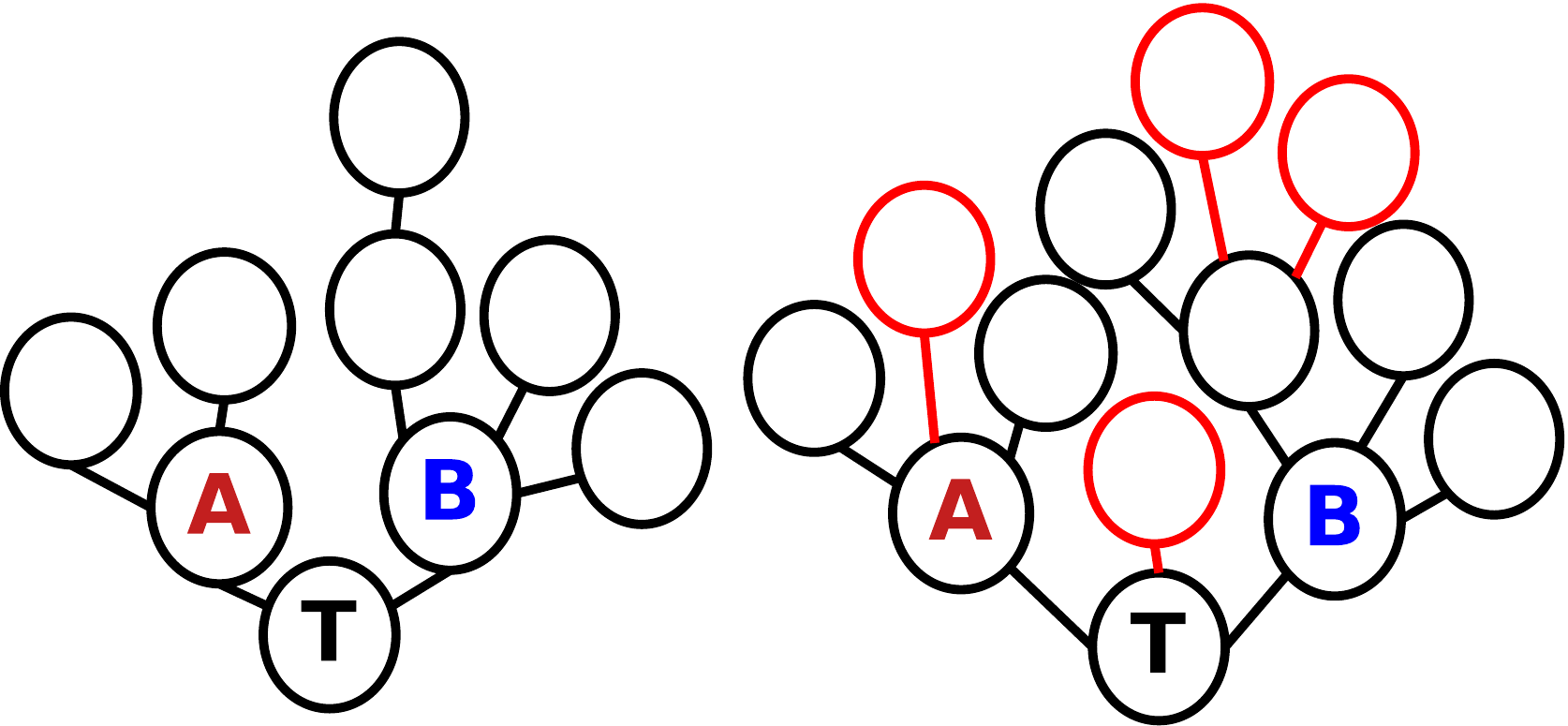}\caption{\label{fig:avap}$\alpha=1$. The tree on the on the right is obtained
from one on the left by adding the branch highlighted in red. The
equation (\ref{eq:alphaone}) implies that, by giving more children
to non extremal branches of height $<H_{max}$, we will always increase
$\mathcal{B}$. In other words, if $H_{max}>2$ then the tree on the
right has a bigger $\mathcal{B}$ then the one on the left.}
\end{figure}

Now, we prove the second result: let us consider a tree. We consider
a branch such that its height is strictly less than $H_{max}$ and
has a number of children $n_{c}$ such that $1<n_{c}<N_{max}$ (so
we precisely pick a branch that \textit{is not} an extremity). If
we build a tree identical to this first one except that the aforementioned
branch is given new children until it has $N_{max}$ children (with
the added children being extremities (i.e. have no child) ) then the
second tree will have a higher $\mathcal{B}$ than the first one.
An illustration of this statement is in Fig.\ (\ref{fig:avap}).
The second step is to take, this time, an extremity and to give it
$N$ children which will become the new extremities, then we can look
at how $\mathcal{B}$ changes. Since we now have $N-1$ more extremities,
the production increases by $(N-1)p_{0}$ but the maintenance cost
increases by $((N-1)(h+1)+N)m_{0}$. With a bit of algebra we can
easily determine the values of $h$ for which the increase in production
compensates the increase in maintenance: if the height of the extremity
we are giving new children is equal or lower than $h_{0}=\left[\frac{p_{0}}{m_{0}}-\frac{N}{N-1}\right] -1$,
then the contribution to $\mathcal{B}$ is positive. It is negative
otherwise, and leads to 

\begin{equation}
H_{0}=\left[ \frac{p_{0}}{m_{0}}-\frac{N_{max}}{N_{max}-1}\right] \label{eq:opti}
\end{equation}

By combining the two previous remarks, we conclude that, below a certain
height, the tree with the highest $\mathcal{B}$ is the tree with
$N_{max}^{H}$ extremities and, beyond this height, any adding new
children to a extremity lowers $\mathcal{B}$.

\section{Generalization for $0<\alpha<1$ and infinite trees\label{sec:Generalization-for-}}

To sum up, the two important facts about $\alpha=1$ we have established
are: no infinitely growing tree is viable and creating as much children
as possible yield the most productive trees as long as the height
of the tree is under a certain height $H_{0}$.

Now, for $0<\alpha<1$, we may ask whether arbitrarily tall trees
that are viable exist. In other words, we want to know whether, for
each height $H$, there exists one tree such that $\mathcal{B}>0$
and what constraints on $p_{0}$ and $m_{0}$ are needed. Obviously,
calculating $\mathcal{B}$ for all possible trees of height $H$ may
not be the best strategy. However, inspired from the results of the
$\alpha=1$ case, we can first limit our search to trees of height
$H$ with $N_{max}^{H}$ extremities (trees maximizing its number
of branches). To illustrate how $\mathcal{B}$ can be calculated,
let us take the example$N_{max}=3$ and $H=3$ then we can easily
count that the trunk has a volume of 27, its 3 children have a volume
of 9, its 9 grandchildren have a volume of 3 and, at last, there are
27 extremities. Therefore, the total maintenance is: 
\begin{equation}
\begin{array}{ccc}
\mathcal{M} & = & m_{0}(27^{\alpha}+3\times9^{\alpha}+9\times3^{\alpha}+27\times1^{\alpha})\\
 & = & m_{0}{\displaystyle \sum_{i=0}^{3}}3^{i}\times3^{(3-i)\alpha}\\
\mathcal{M} & = & 3^{3\alpha}m_{0}{\displaystyle \sum_{i=0}^{3}}3^{(1-\alpha)i}
\end{array}\label{eq:intermed}
\end{equation}

Using the same kind of reasoning for a more general $N_{max}$ (that
we will note as $N$ from now on) and any $H$, we can deduce the
Eq.\textvisiblespace{}(\ref{eq:c_H}):

\begin{equation}
\begin{array}{cccc}
\mathcal{M}_{H} & = & N^{\alpha H}m_{0}{\displaystyle \sum_{i=0}^{H}}N^{(1-\alpha)i}\\
\\
 & = & N^{H}m_{0}\frac{N^{1-\alpha}-N^{-(1-\alpha)H}}{N^{1-\alpha}-1} & \mathrm{,\:if\:}N>1\mathrm{\:and\:\alpha\neq1}
\end{array}\label{eq:c_H}
\end{equation}

From there we have an expression for $\mathcal{B}_{H}$ when $N>1$
and $\alpha\neq1$ :

\begin{equation}
\mathcal{B}_{H}=N^{H}\left(p_{0}-m_{0}\frac{N^{1-\alpha}-N^{-(1-\alpha)H}}{N^{1-\alpha}-1}\right)\label{eq:P_h-C_h}
\end{equation}

From this expression we can see that if $0<\alpha<1$, then $p_{0}-m_{0}\frac{N^{1-\alpha}-N^{-(1-\alpha)H}}{N^{1-\alpha}-1}\geq p_{0}-m_{0}\frac{N^{1-\alpha}}{N^{1-\alpha}-1}$
for every $H$ therefore, for a given $N$ and $\alpha$, the value
of $p_{0}/m_{0}$ determines whether $\mathcal{B}_{H}$ will eventually
become negative. So, we have the following two cases:
\begin{equation}
\begin{array}{c}
p_{critical}=\frac{N^{1-\alpha}}{N^{1-\alpha}-1}\\
p_{0}/m_{0}\geq p_{critical}\textrm{ }\Rightarrow\forall H,\textrm{ }\mathcal{B}_{H}\geq0\\
p_{0}/m_{0}<p_{critical}\textrm{ }\Rightarrow\exists H_{f},\textrm{ }\forall H\geq H_{f},\textrm{ }\mathcal{B}_{H}<0
\end{array}\label{eq:cases1}
\end{equation}

In the first case, $\mathcal{B}_{H}\rightarrow\infty$ and, in the
second case, $\mathcal{B}_{H}$ becomes strictly decreasing after
a certain $H$, then $\mathcal{B}_{H}\rightarrow-\infty$. So, a sufficient
condition for the existence of viable trees (i.e. $\mathcal{B}>0$)
at any height is $p_{0}/m_{0}\geq p_{critical}$.

\begin{figure}
\hfill{}\includegraphics[width=250bp]{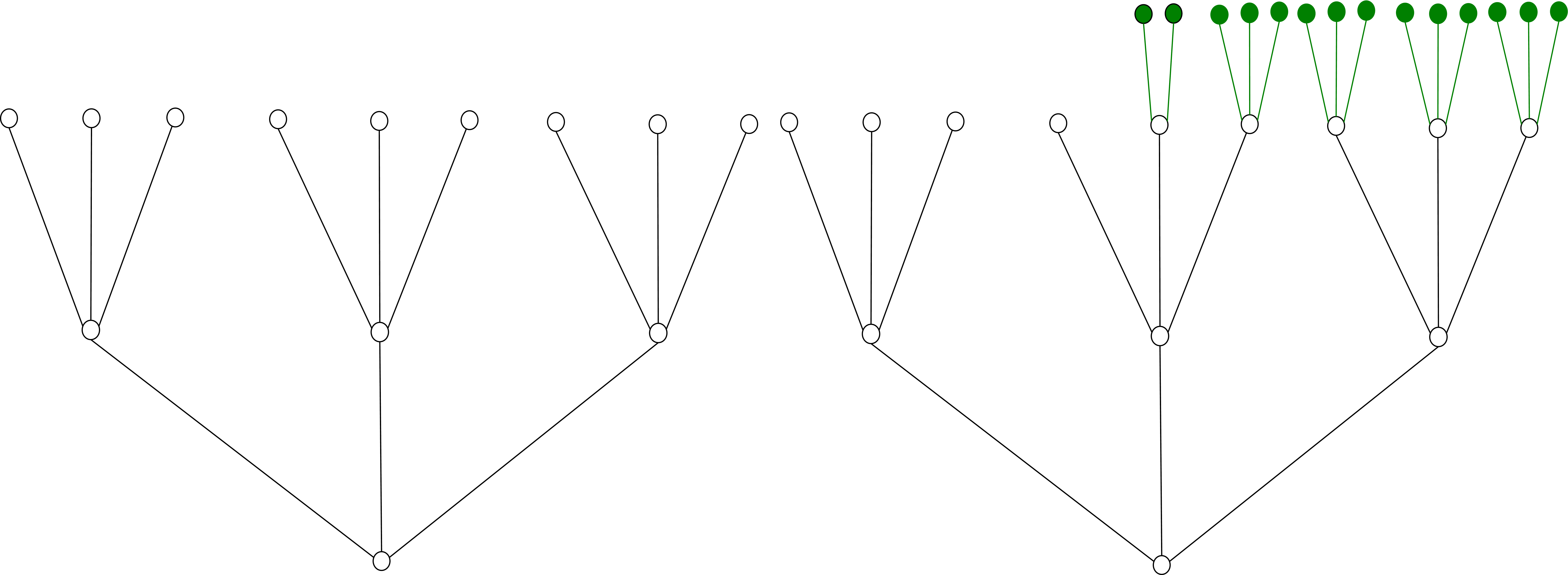}\hfill{}

\caption{\label{fig:Here,-the-maximal}Here, $N,$the maximal number of children
a branch may have, is equal to 3. If we want to construct a tree with
$m$ extremities (in this example $m=18$), then one way it can be
done is by starting from tree of height $H$ (in our example, $H=3)$
maximizing its number of children then we give children to consecutive
extremities until we reach $m$, like in the figure above. This way
of constructing a tree with $m$ extremities minimizes the total amount
of volume the tree needs to have (assuming Leonardo's rule applies).}
\end{figure}

This condition is also necessary, but we need to go through multiple
arguments to reach this conclusion. A first observation to make is
that branches have no benefit having only one child since only having
one do not increase the number of extremities while increasing the
maintenance. So, we will only discuss trees for which every branches,
except the extremities, have at least 2 children. With this kind of
trees, for any $m\in\mathbb{N}$, we can find $H_{m}\in\mathbb{N}$
such that any trees of height $H_{m}$ or higher have $m$ or more
extremities.

We make a second observation: let $m\in\mathbb{N}$, we can find $H\in\mathbb{N}$
such that $N^{H}\leq m<N^{H+1}$. If we construct a tree with $m$
extremities by starting from the tree of height $H$ that has $N^{H}$
extremities, then adding the remaining $r_{0}=m-N^{H}$ new extremities
in the way described in Fig.\ (\ref{fig:Here,-the-maximal}), this
tree has a maintenance cost lower or equal than any other trees with
$m$ extremities. Let us verify this last statement. First, it is
obvious we there can not exist a tree with $m$ extremities that is
strictly shorter than the one we just constructed because no branch
may have more than $N$ children. Second, let us consider strictly
taller trees with $m$ extremities: they would have more total volume
(since we recall the formula for the total volume is $\sum_{h}(h+1)E_{h}$)
and share this total volume among more branches, therefore, using
the inequality $(\sum x)^{\alpha}\leq\sum x^{\alpha}$, we deduce
their maintenance cost must be higher. Third, now we simply need to
compare trees which height is $H+1$: they all look like our tree
except that the $r_{0}$ remaining extremities may be split differently.
Our splitting minimizes maintenance costs because $v^{\alpha}+(v+x_{1}+x_{2})^{\alpha}\leq(v+x_{1})^{\alpha}+(v+x_{2})^{\alpha}$
which means it is better to pack the $r_{0}$ branches together in
one side.

A third and last observation has to be made. We mentioned that when
the condition is not respected (i.e. $p_{0}/m_{0}<p_{critical}$),
$(\mathcal{B}_{H})_{H}$ is strictly decreasing and negative after
some $H_{f}$. We can establish that if $H\geq H_{f}$, then the tree
with $m$ (such that $N^{H}\leq m<N^{H+1}$) extremities described
earlier (Fig.\ (\ref{fig:Here,-the-maximal})) must have a $\mathcal{B}$
inferior to $\mathcal{B}_{H}$ (therefore negative too). Indeed, $\mathcal{B}$
not being inferior to $\mathcal{B}_{H}$ iscontradictory: we name
again $r_{0}=m-N^{H}$ and assume $r_{0}=N-1$ then conclude that
if the resulting $\mathcal{B}$ is superior to $\mathcal{B}_{H}$,
then $\mathcal{B}_{H+1}\geq\mathcal{B}_{H}$ because in this case
adding one (or more) group of $N-1$ extremities will increase $\mathcal{B}$
even more. Consequently, the case $r_{0}=N-1$ must lowers $\mathcal{B}$
which we can then use to prove the case $r_{0}=k(N-1)$ also lowers
it, and after this we deduce it for a more general $r_{0}$.

Finally, using the first observation, we can find $H_{1}$ such that
every trees of height higher than $H_{1}$ have at least $N^{H_{0}}$
extremities. Then, for each of these tree, using the second observation,
we can create a tree, similar to the one in Fig.\ (\ref{fig:Here,-the-maximal})
with the same number of extremities but with a lower $\mathcal{B}$.
But, this last tree has a height $H\geq H_{0}$ so its $\mathcal{B}$
is negative, using the third observation. Thus, when $p_{0}/m_{0}<p_{critical}$,
arbitrarily tall and viable trees do not exist.

Now, let us go back to Eq.\ (\ref{eq:P_h-C_h}) and contextualize
it. Until now, we looked at the static cost of the branches. By doing
so we ignored an important aspect of our dynamical tree: branches
need to actually spend resource to grow into the volume of Leonardo's
rule; each time a branch increases in volume it consumes an amount
proportional to this volume increase (Sec.\ \ref{sec:Example-of-implementation}:
$C_{v}$ being the proportionality coefficient). Let us look at the
cost induced by this process when we go from the tree of height $H-1$
and $N^{H-1}$ extremities to the tree of height $H$ with $N^{H}$
extremities.

\begin{equation}
\begin{array}{ccc}
\mathcal{C}_{vol(H-1\rightarrow H)} & = & C_{v}\textrm{(number of additional extremities)}H\\
 & = & N^{H}C_{v}(1-N^{-1})H
\end{array}\label{eq:C_vol}
\end{equation}

If we can compare $\mathcal{C}_{vol(H-1\rightarrow H)}$ to $\mathcal{B}_{H}$,
it is $\mathcal{C}_{vol(H-1\rightarrow H)}$ that dominates for large
$H$ ($\mathcal{B}-\mathcal{C}_{vol}$ tends to $-\infty$ as $H$
grows). However unlike a maintenance cost, this creation cost is only
paid once each time the tree grows, so a tree that takes time to accumulate
a reserve between each growth spurt may grow arbitrarily tall. But
the amount of time waiting to accumulate a reserve will increase as
$\mathcal{O}(H)$. Simply put, if $C_{v}\neq0$ then our tree will
not be able to grow infinitely tall unless it is ``intelligent''
enough to slow down its growth, and even then we predict a ``rate
of slow down'' that is exponential. It may echo the fact real trees
have their growth slowing down with size\cite{Mencuccini2005}.

\section{Simulation of symmetric trees\label{sec:Simulation-of-symmetric}}

\subsection{Encoding the trees in sequences of integer}

Up until now, all our theoretical results are independent of the dynamical
rules we implement: they all are simple arguments on topology and
economic costs. Let us go back to the dynamical system described in
section \ref{sec:Example-of-implementation}, and we consider the
case where all the branches has the same strategies during flux up
(the phase in which the flow go from the trunk to the extremities):
at the end of each generation, the extremities generate as much children
as possible with the resource they possess. As explained in Sec.~\ref{sec:Example-of-implementation}:
without asymmetry or randomness, the tree will be ``symmetric'',
meaning that all branches located at the same height will have the
same number of children and brothers, the same volume etc. The trunk
will have $u_{0}\leq N$ number of children. By symmetry, these $u_{0}$
children will all have the same number of offspring, noted $u_{1}\leq N$,
etc. So, for a tree of height $H$, we can construct the sequence
$u_{0}$,$u_{1}$,...,$u_{H-1}$ where $u_{i}\leq N$. A tree composed
of only a trunk is represented by an empty sequence. Conversely such
a sequence will define a symmetric tree of height $H$.

Hence we can see the growth of our simulated symmetric trees as a
finite sequence of $u_{i}\leq N$ that evolves dynamically. While
the ''form'' of a symmetric tree of height $H$ is characterized
by a sequence $u_{0}$,...,$u_{H-1}$, the tree itself is also defined
through how the volume and the reserve is allocated among the branches.
Since every branches located at the same height have the same volume
and reserve then the full characterization of a tree of total height
$H$ is given by 3 finite sequences: $(u_{i})_{0\leq i\leq H-1}$,
$(v_{i})_{0\leq i\leq H}$ representing the volume $v_{i}$ of the
branches located at height $i$ and $(r_{i})_{0\leq i\leq H}$ giving
the amount of reserve $r_{i}$ stored in the branches located at height
$i$. With our symmetrical dynamics the sequence $(u_{i})_{0\leq i\leq H-1}$
may shrink (branches of height superior to some $h$ die) or may grow
(new integers $u_{H}$, $u_{H+1}$ etc added) but the values $u_{i}$
themselves do not change because if a branch of height $h+1$ dies,
those of the same height also all die by symmetry, therefore $u_{h}$
would simply disappear from the sequence as well as all $u_{i>h}$.

The initial condition for all our simulation is $(u_{i})=\varnothing$
(symbol for empty sequence), $(v_{i})=(1)$ and $(r_{i})=(0)$ (i.e.
we always start with a tree formed only by a trunk of volume 1 with
no initial reserve). Given how, we kill every branch that has a volume
less than what Leonardo's rule would predict then, as already mentioned,
the only time some $v_{i}$ do not follow Leonardo's rule is when
a tree got some branches dead (so that the survivors will have bigger
volume than predicted by Leonardo's rule). So if we interest ourselves
with infinitely and steadily growing trees the sequence $(v_{i})_{i}$
is unneeded since it matches Leonardo's rule perfectly. Though, $C_{v}=0$
is a requirement if we want trees growing ever steadily.

\subsection{Evolution of $(u_{i})_{i}$ for infinitely and steadily growing trees}

We define a ``steadily'' growing tree as a tree such that at the
end of generation number $i$, the height of the tree is also $i$
(i.e. the tree never shrinks and never stagnates). We have stated
at the end of Sec.\ \ref{sec:Generalization-for-} that unless $C_{v}=0$
an infinitely and ``steadily'' growing tree is impossible, so from
now on $C_{v}$ is taken as 0 (i.e. growing in diameter/volume do
not cost any resource to the branches). If we note $B_{H}=u_{0}\times u_{1}\times...\times u_{H-1}$
and define $B_{0}=1$ then the $B_{i}$ represent the total number
of branches located at the height $i$. If we want to write an expression
for $u_{H+1}$ or simulate the system, we first need to place ourselves
in a specific version of the system described in Sec.~\ref{sec:Example-of-implementation}.
For the specific system we study: during ``flux down'', the children
give everything to their parent which means that at the beginning
of ``flux up'' the trunk start with the total production. Then,
during ``flux up'', the parents only pay their maintenance, are
eliminated if they can not pay and then grow in volume (which is free
since $C_{v}=0$) but do not keep anything more in reserve and distribute
equally to their children the remaining flux. Finally when the flux
reaches the extremities, they, after paying their maintenance, use
as much resource as possible to create as many children as they can
(the cost of creation for each child being an integer $\mathcal{C}$).
With this choice of evolution rule, the system is greatly simplified
because, right before the extremities have to create new children
and at the end of all the maintenance payment, the resource the extremities
have is equal to the total amount produced (by the extremities at
the beginning of ``flux down'') minus the total maintenance cost
of the tree plus the leftover resource that could not be transformed
into children the turn before, noted $R_{i}$. So since we know the
resource the extremities have right before creating children, we can
predict the number of children each extremity will spawn, $u_{H+1}$:

\begin{equation}
\begin{array}{c}
u_{H+1}=\left[ \frac{B_{H+1}\left({\displaystyle p_{0}-\sum_{i=0}^{H+1}\left(\frac{B_{i}}{B_{H+1}}\right)^{1-\alpha}m_{0}}\right)+R_{H+1}}{\mathcal{C}B_{H+1}}\right] \end{array}\label{eq:u_h+1_tot}
\end{equation}

Of course, if there is a limit $N_{max}<\infty$, we take the minimum
between the expression above and $N_{max}$. 

$R_{H+1}$ is the unused leftover during the previous cycle in other
words: 
\begin{equation}
R_{H+1}=B_{H}\left({\displaystyle p_{0}-\sum_{i=0}^{H}\left(\frac{B_{i}}{B_{H}}\right)^{1-\alpha}m_{0}}\right)+R_{H}-B_{H+1}\mathcal{C}\label{eq:reserve}
\end{equation}
The big first term in the numerator of Eq.\ (\ref{eq:u_h+1_tot})
is the total production minus the total maintenance. And at the denominator,
we take into account the fact the $B_{H+1}$ extremities share this
total resource and $\mathcal{C}$ is the cost of a child. This formula
only works as long as the tree grows ``steadily'', meaning that
if along the flux up some branches could not pay maintenance or if
the extremities do not have enough to create a child ( $u_{H+1}=0$
), then (\ref{eq:u_h+1_tot}) and (\ref{eq:reserve}) stop being predictive.
Conversely, as long as all the $u_{H+1}$ give strictly positive integer,
we can be sure no branches died along ``flux up'' (otherwise the
numerator in (\ref{eq:u_h+1_tot}) would be negative and so would
$u_{H+1}$). In other words, the moment $u_{H+1}\leq0$ is the moment
the tree has stopped growing ``steadily'' and we note $H_{f}$ the
last height/generation (\ref{eq:u_h+1_tot}) is valid. $H_{f}$ may
or may not be infinite but we are particularly interested in the infinite
case.

\subsubsection{The case $R_{i}=0$: the tree does not keep its leftover resource }

Now we put the term $R_{H+1}$ to zero and since $u_{H}=B_{H+1}/B_{H}$,
we obtain a simple expression to get the successive $B_{H+1}$.
\begin{equation}
B_{H+1}=B_{H}\cdot\mathbf{min}\left(\left[ \frac{p_{0}-\sum_{i=0}^{H}\left(\frac{B_{i}}{B_{H}}\right)^{1-\alpha}m_{0}}{\mathcal{C}}\right] \:,\:N_{max}\right)\label{eq:B_H1}
\end{equation}
To simplify things we introduce $W_{n}={\displaystyle \sum_{i=0}^{n}(B_{i}/B_{n})^{1-\alpha}}$.
\[
W_{n+1}=\left(\frac{B_{n\text{+1}}}{B_{n+1}}\right)^{1-\alpha}+\sum_{i=0}^{n}\left(\frac{B_{i}}{B_{n+1}}\right)^{1-\alpha}=1+\sum_{i=0}^{n}\left(\frac{B_{i}}{B_{n+1}}\right)^{1-\alpha}\:.
\]
Then we replace the $B_{n+1}$ term at the denominator by the expression
in (\ref{eq:B_H1}) which replaces $B_{n+1}$ by $B_{n}$ multiplied
by a factor. This term $B_{n}$ allows us to recover a $W_{n}$, and
we end up with a ``simple'' recurrence.

\begin{equation}
\begin{array}{c}
W_{n+1}=1+W_{n}\dfrac{1}{\left(\mathbf{min}\left(\left\lfloor a_{0}-W_{n}\cdot b_{0}\right\rfloor \:,\:N_{max}\right)\right)^{1-\alpha}}\\
\\
\textrm{Where }a_{0}=p_{0}/\mathcal{C}\textrm{ and }b_{0}=m_{0}/\mathcal{C}\\
\\
\textrm{with }W_{0}=1
\end{array}\label{eq:W_n+1}
\end{equation}

Just as with Eq.~(\ref{eq:u_h+1_tot}), this expression is valid
only until $n=H_{f}$ which corresponds to the first integer that
yields $\left\lfloor a_{0}-W_{H_{f}}\cdot b_{0}\right\rfloor \leq0$.
Since we are interested in trees that can grow infinitely we limit
our study to $0<\alpha<1$, and in this case $H_{f}$ may be infinite
for some values of $a_{0}$ and $b_{0}$. $H_{f}=\infty$ is equivalent
to $W_{n}$ (equivalently $B_{n}$) being defined for any natural
number $n$. And if these two sequences are defined for every $n$
we may want to look if they converge or not. First we need to establish
when do we get an infinite sequence ( $H_{f}=\infty$ ). We also remind
$a_{0}$, $b_{0}$$>0$. $p_{0}>m_{0}$ is also assumed because otherwise
our tree will stop growing at $H_{f}=1$.

We define the function $f_{0}$ such that Eq.~(\ref{eq:W_n+1}) simply
becomes a discrete dynamical system $W_{n+1}=f_{0}(W_{n})$ and $f_{1}(x)=1+x/\left(a_{0}-x\cdot b_{0}\right)^{1-\alpha}$,
that we can study with classical tools. In fact $f_{1}$ is a simplified
and a more easy-to-study version of $f_{\text{0}}$. Their behavior
with respect to $\mathbb{I}:x\rightarrow x$ provides us with the
behavior of $(W_{n})$.
\begin{figure}
\begin{centering}
\includegraphics[width=7cm]{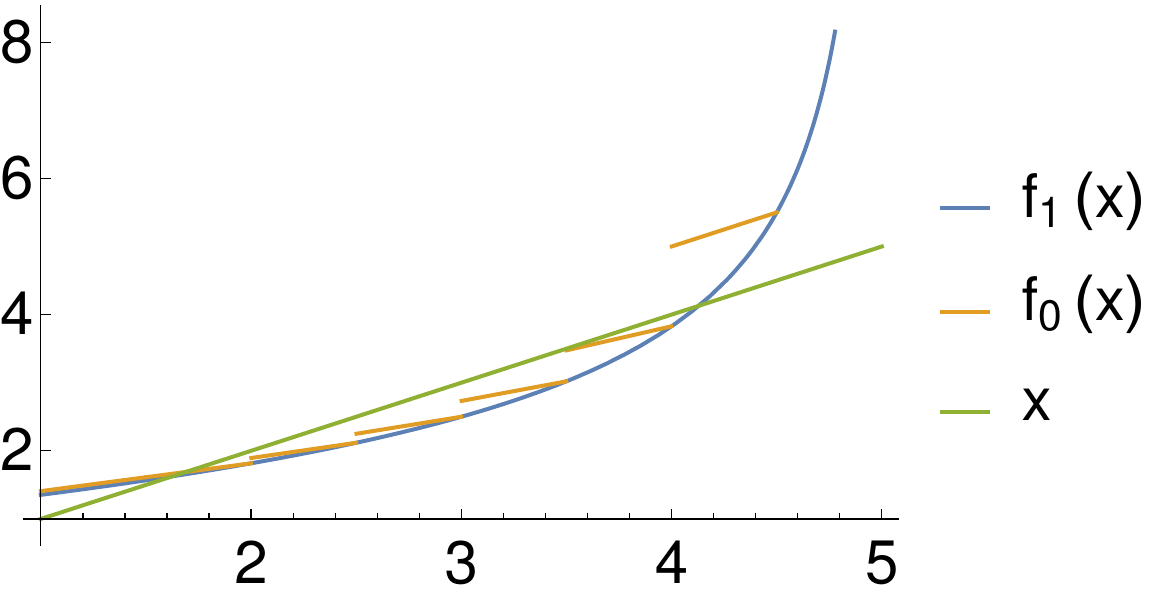}
\par\end{centering}
\caption{\label{fig:3}Plot of $f_{0}$, $f_{1}$ and $\mathbb{I}$ for $N_{c_{max}}=6$
and $\alpha=0.5$. For $0<\alpha<1$, the $f$ functions are monotonous
increasing and $f_{1}$ intersects twice the diagonal line. Here,
$a_{0}=10$ and $b_{0}=2$.}
\end{figure}
 The figure (\ref{fig:3}) provides us with the behavior of these
functions with respect to one another. Before moving to study this
dynamical system, we point out a few things about $f_{0}$, $f_{1}$
and $W_{n}$:
\begin{itemize}
\item The domains of definition of our functions are $\mathcal{D}(f_{1})=(-\infty,\textrm{ }a_{0}/b_{0})$
and $\mathcal{D}(f_{0})=(-\infty,\textrm{ }(a_{0}-1)/b_{0}]$. The
functions being positive and the initial condition being $W_{0}=1$,
only $\mathbb{R}^{+}\cap\mathcal{D}(f_{1})$ and $\mathbb{R}^{+}\cap\mathcal{D}(f_{0})$
interest us. The sequence $(W_{n})_{n}$ stops when $f_{0}(W_{n})$
escapes the domain of definition of $f_{0}$ and the term $H_{f}$
is the last defined one.
\item Given $a_{0},$ $b_{0}>0$ and $0<\alpha<1$, both $f_{0}$ and $f_{1}$
are strictly increasing in the domains we are interested in.
\end{itemize}
Then we can show that even for a non-continuous function like $f_{0}$,
we end up with an attractor which is a simple fixed point. (see Appendix~\ref{sec:Appendix}
for details) this means that the sequence $W_{n}$converge in most
cases, it is indeed possible that $f_{0}$ is not below the line $y=x$.
\begin{figure}
\begin{centering}
\includegraphics[width=7cm]{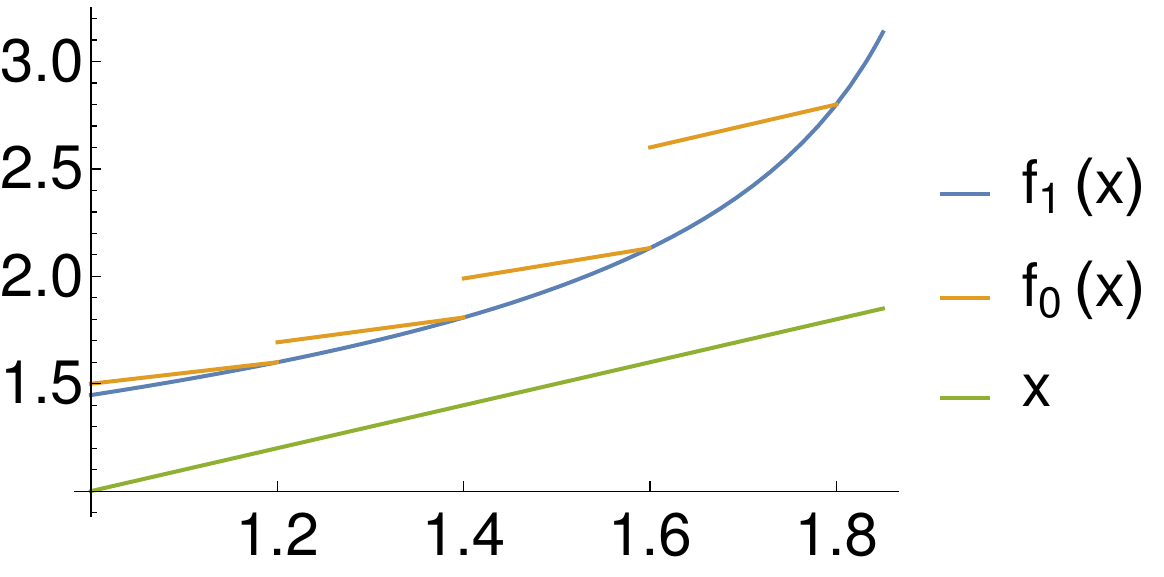}
\par\end{centering}
\caption{\label{fig:Using}Using different parameters for $a_{0}$ and $b_{0}$
than in figure (\ref{fig:3}), we can elevate the graphs of $f_{0,1}$
above the line $y=x$. In such case, we are sure that $(W_{n})$ is
finite since one $W_{i}$ will go beyond the domain of definition
of $f_{0,1}$.}
\end{figure}

Now let us suppose we do have a fixed point. We call this fixed point/
limit $W^{*}$. But 
\begin{equation}
 u_{n}=\mathbf{min}\left(\left[ a_{0}-W_{n}b_{0}\right] \:,\:N_{max}\right)\:,
\end{equation}
so $(u_{n})_{n}$ converges too. If we assume $N_{max}>\left\lfloor a_{0}-W^{*}b_{0}\right\rfloor $
then $u^{*}=\lfloor a_{0}-W^{*}b_{0}\rfloor$. So if we fix $a_{0}$
and $b_{0}$, the limits of both $(W_{n})$ and $(u_{n})_{n}$ will
be tied together. The conclusion, which is observable through the
simulations, is that when we negate the reserve $R=0$ then infinite
trees give sequences $(u_{i})$ that ends with an infinite succession
of $u^{*}$.

In conclusion, with the dynamics and strategy we described (there
is no external threat and the branches are acting brainlessly), infinite
trees only appear for some values of the parameters and they end up,
after a certain time, growing in a very regular way: the number of
extremities is multiplied by $u^{*}$ each generation and we end up
with a self-similar tree.

\subsubsection{Case with the reserve on: back to equation (\ref{eq:u_h+1_tot}).}

\begin{figure}
\hfill{}\includegraphics[width=160bp]{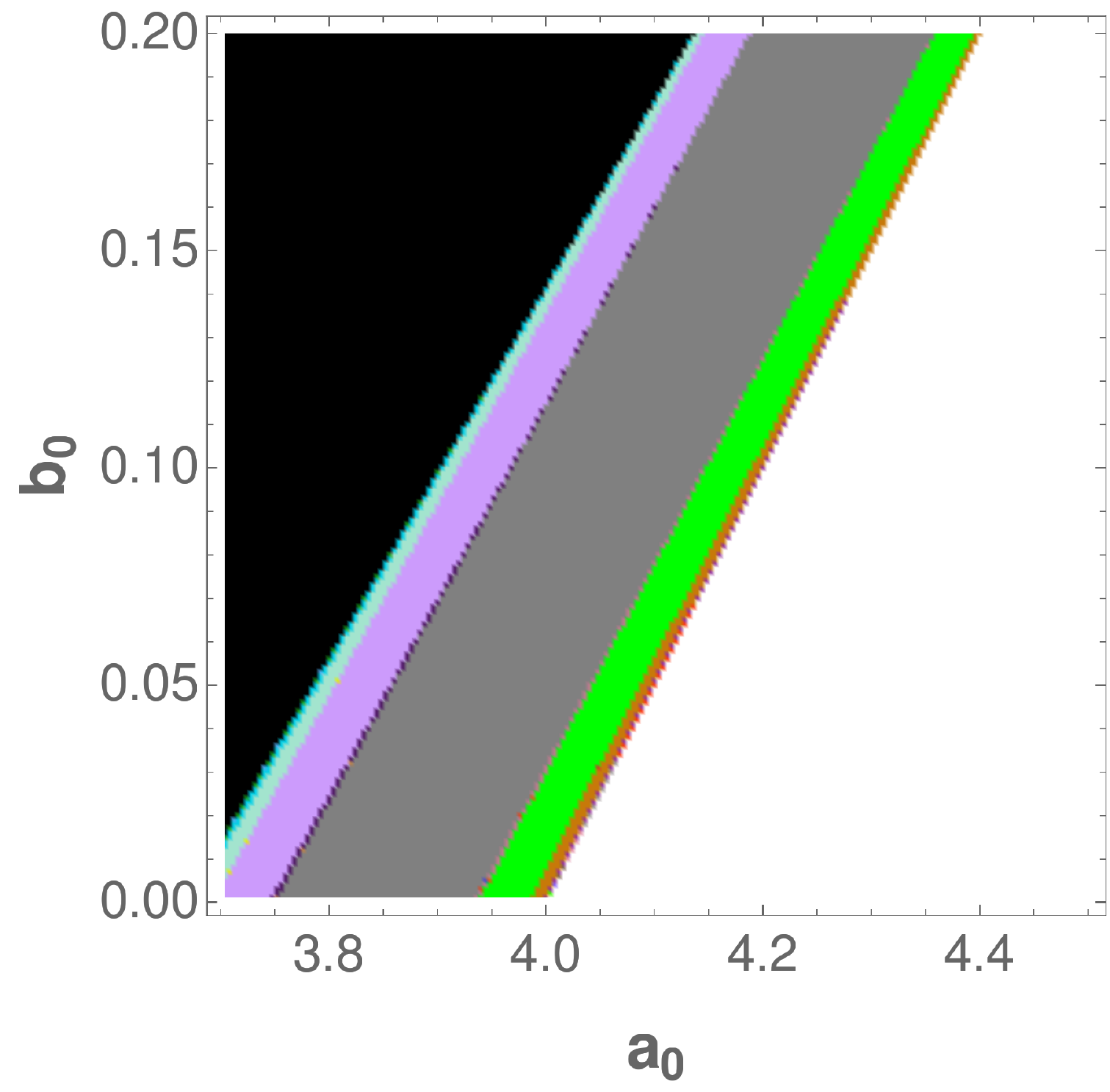}\hfill{}

\caption{\label{fig:last} The sequences $(u_{i})_{i}$ become periodic after
a few terms. Here, for $\alpha=0.5$, we noted down the period we
get for different values of $a_{0}$ and $b_{0}$ and we found 36
different periods. Each of them is encoded by one color. The period
on the top-left corner encoded in black is the period composed only
of 3, noted (3), while the one on the bottom-right corner encoded
in white is the period (4). We can see that between (3) and (4), we
go through a multitude of bands of different color. These periods
are a succession of $3$ and $4$ but some have a bigger proportion
of 4 or 3: the one near the (4) region has more 4 and vice versa.
The large band in gray on the middle represents the period (3, 4)
and it separates regions where the proportion of 4 is larger from
the ones with more 3. There appears to be some self-similar patterns
at the frontier of two ``large'' domains. There, we made a graph
around $a_{0}=4$ but if we make one around an other integer $a_{0}=N$,
the main difference will be that the periods are composed of the integers
$N$ and $N-1$ but other than that we will see the same pattern of
bands.}
\end{figure}

First we can make the following remark: even with the reserve turned
on, the results about the necessary conditions to get an infinitely
growing tree described earlier should, at least to some extent, hold
true. However with the addition of the reserve, we observe (from numerical
calculations) that instead of having a sequence $(u_{i})_{i}$ that
ends with an infinite succession of $u^{*}$, we may sometimes have
a periodicity: the sequence will end up oscillating between $u_{1}^{*}=\lfloor a_{0}-W^{*}b_{0}\rfloor$
and $u_{2}^{*}=\lceil a_{0}-W^{*}b_{0}\rceil$. There does not seem
to be limits for the length of the periods we find. But, the period
we get, vary progressively in function of the parameters $a_{0}$
and $b_{0}$ (Fig.\ \ref{fig:last}). So while we have a less repetitive
growth than in the case $R=0$, the tree grows in a very regular manner
and do not seem to change erratically its growth pattern when we vary
slowly the parameter values.

\section{Conclusion and perspectives\label{sec:Conclusion}}

\begin{figure}
\centering{}\includegraphics[width=150bp]{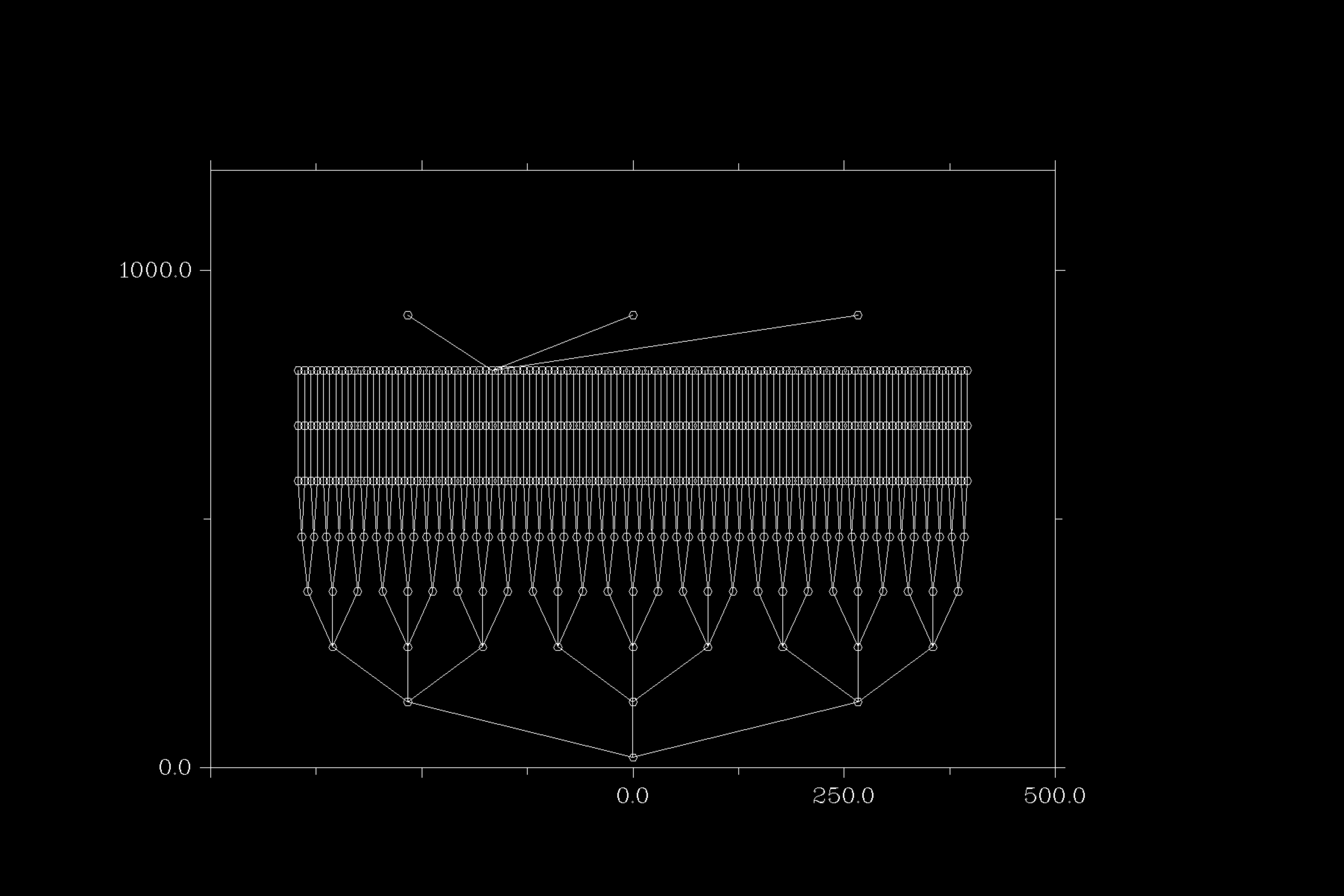}\\
\includegraphics[width=150bp]{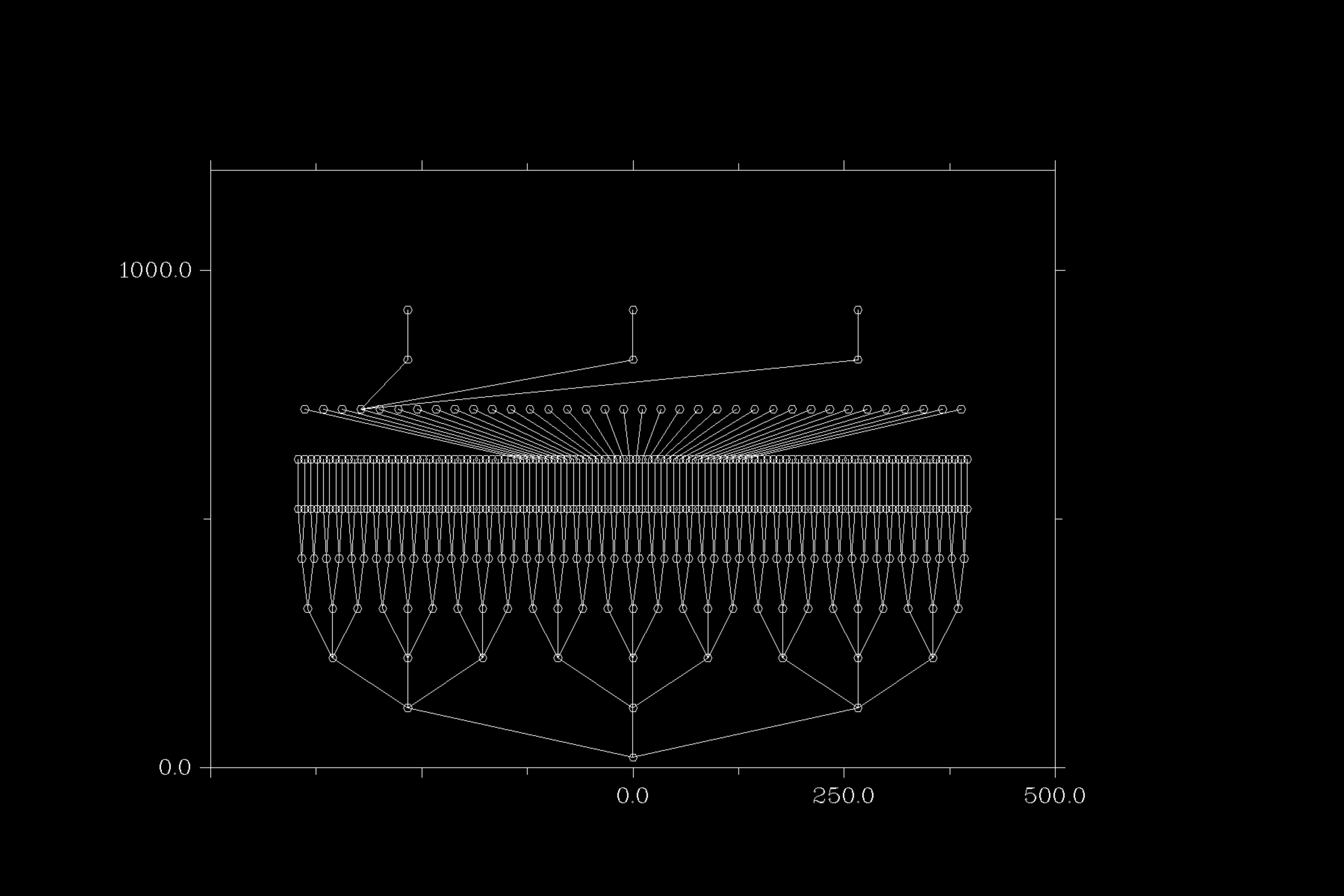}\\
\includegraphics[width=150bp]{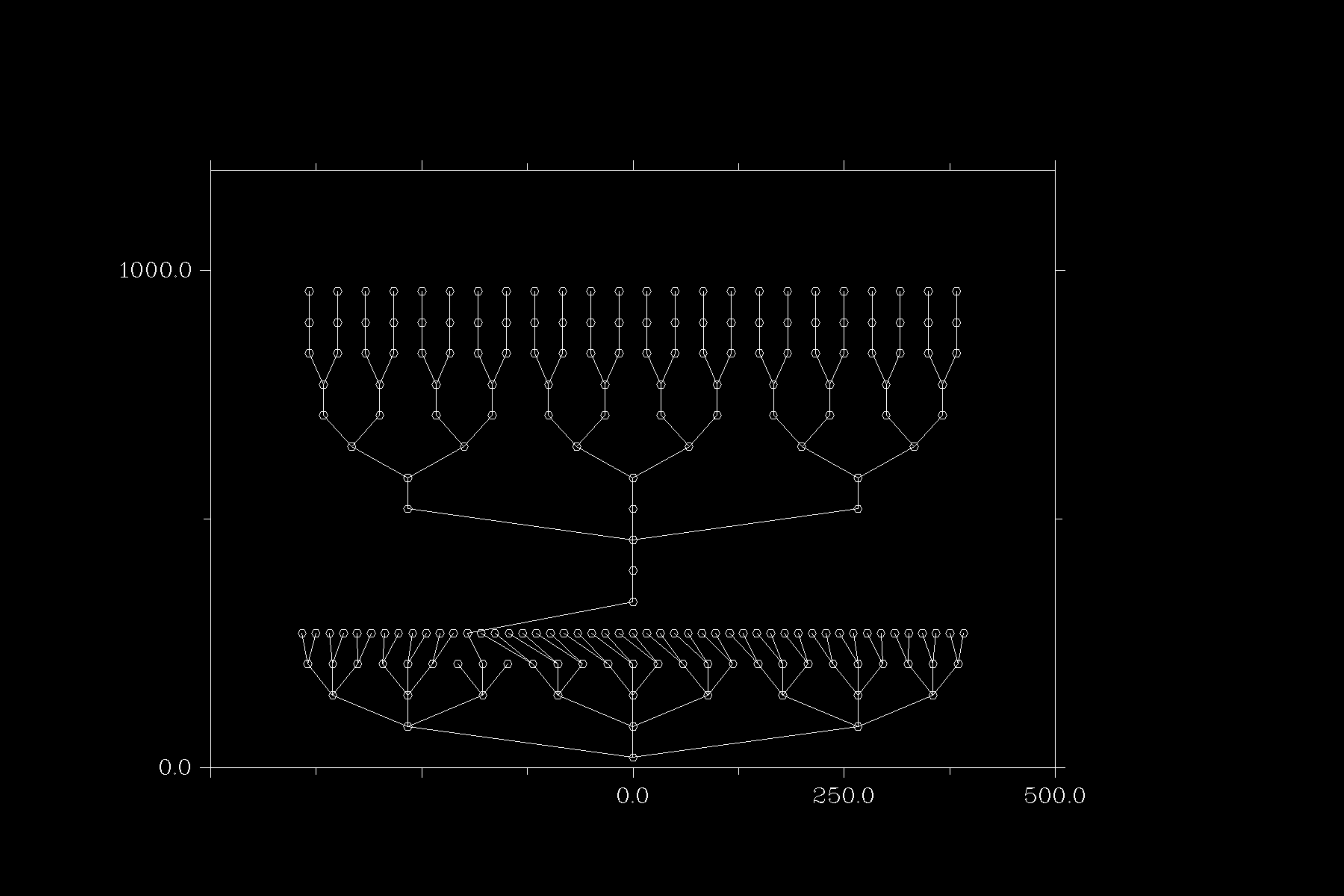}\caption{\label{fig:last_fig}A representation of three trees. The node at
the bottom is the trunk. An eternal tree is perturbed adding 3 extremal
branches (Top). Consequences of redistribution rules are shown in
the middle and the bottom figure. The ``rewarding'' strategy ( $r_{1}=0.3$
and $r_{2}=1$) leads to an eternal tree (middle one) while the ``altruist''
strategy ($r_{1}=2$ and $r_{2}=1$) leads to a tree that ends up
dying (bottom figure) . The values for the other parameters are: $p_{0}=400$,
$m_{0}=50$, $C_{v}=10$, $\mathcal{C}=70$ and $\alpha=1$. To be
more precise the tree with rewarding strategy eternally oscillates
between forms similar to the ones displayed in the top and middle
figures, while the bottom tree dies shortly after, being unable to
support its structure, this would lead to indicate a survival of the
fittest strategy would be best in the case of this individual tree.}
\end{figure}
In this paper, we have presented a set of basic rules describing a
group of bio-inspired dynamical systems focused on resource distribution
and allocation. By completing these basic rules with specific evolution
rules, we construct a growing tree-like system we can simulate and
study. However, even with only the basic rules and no evolution rules,
we have shown it was possible to get a few results about the possible
topologies of a tree-network that possesses sources at its extremities
and have a maintenance cost for each node which increases the closer
the node is to the root (Leonardo's rule). Then, we specified the
evolution rules and studied the resulting dynamical system. By studying
a very simple version of the system, we could established a few theoretical
and numerical results that can be useful to lay some ground for future
works. We emphasize as well that during our study we initially spend
a lot of time scanning parameter space for the case $\alpha=1$, in
this situation, as already discussed, most trees end up dying, however
for some carefully chosen parameters we can end up with what would
be an like an unstable fixed point, a tree that simply stops growing
and lives forever. Nothing really interesting came out of this thorough
numeric study which is why we did not present it in this paper, however
these unstable tree allowed us to test the consequence of how the
redistribution of resources among children affects the tree. The results
are displayed in Fig.~\ref{fig:last_fig}, we start from a perturbation
of an unstable tree by adding three extra branches, this will lead
to a depletion of the reserves and a death of the tree, but as illustrated
we can see that depending on how the redistribution is made, the tree
can be more resilient depending on the redistribution chosen . The
formula that dictates the proportion a branch/parent will distribute
its resource to each child depends on two elements: 1) the need of
the child which is composed of its maintenance cost plus the cost
of growing to reach the required volume (Leonardo's rule) 2) the amount
of resource the child gave to the parent during ``flux down''. More
specifically, if we note $e_{i}$ the need of a child $i$ and $c_{i}$
its contribution during flux down then the formula is $Ze_{i}^{r_{1}}c_{i}^{r_{2}}$.
With $r_{1}$ and $r_{2}$ parameters and $Z$ the normalization constant
${\textstyle \sum_{j}}e_{j}^{r_{1}}c_{j}^{r_{2}}$ (sum over all the
children $j$ of this particular parent), and thus if $r_{1}>r_{2}$
the tree redistribute resources based on needs, while the $r_{1}<r_{2}$
is based more on reward. The results displayed in Fig.~\ref{fig:last_fig}
clearly show that for the considered case, a strategy based on reward
keeps the tree alive, while the one based on needs ends up killing
the tree. This leaves us with the perspective of this work: we dubbed
this paper part 1 kinematics, as no external forces or interaction
between the branches besides redistribution of the resources was taken
into account, so only the self-sustained kinematics of the tree were
taken into account. As a first perspective we want to embed this growth
into real space, which will add occupation constraints on the new
branches and some exclusion rules of available space to grow new offspring.
This will lead to some interactions between the branches and we expect
that the resulting dynamics will be greatly affected by this. Another
aspect of future work, will be to influence the role of the redistribution
parameter and its possible influence on the resilience of the tree
structure as well as its overall shape like its extremities (foliage)
when embedding it in real space. A comparison would be then possible
with the already existing a attempts to obtain realistic looking tree
shape from relatively simple rules \cite{Eloy2017,Duchemin2018}.
Thus, with the embedding in space, we could examine whether our approach
with resource distribution and allocation could yield similar results.
\section*{Ackowledgements}

X.L. thanks Christophe Eloy, for fruitful discussions and contributions especially in the early development of the model,
and its numerical implementation. O.B. and X.L. would like as well to thank Bruno Moulia, Eric Badel and
Andr\'e Lacointe for encouragements and useful suggestions.
The project leading to this publication has received funding
from Excellence Initiative of Aix-Marseille University -
A*MIDEX, a French ?Investissements d?Avenir? programme.
It has been carried out in the framework of the Labex MEC.
We also acknowledge support from the CNRS (Mission pour
l?interdisciplinarité, project ARBRE)

\section*{Appendix A\label{sec:Appendix}}

Let us show that even for a non-continuous function like $f_{0}$,
we end up with an attractor which is a simple fixed point. Indeed,
any real-valued function $g$ strictly increasing defined on $[x,\:y]$
such that $g(x)>x$ and $g(y)<y$, $g$ has a fixed point on $[x,\:y]$.
Likewise with only these constraints on $g$ , if we define $u_{n+1}=g(u_{n})$
with $u_{0}=x$ as initial condition then the sequence is define for
any $n$ and converges. We can apply this result to the function $f_{0}$
and the sequence $(W_{n})_{n}$ with $W_{0}=1$. First, we may want
to set $a_{0}>b_{0}+1$ so that $W_{1}=f_{0}(1)$ is defined. And
if it is defined then it is obvious $f_{0}(1)>1$. Therefore, the
last step is to prove the existence of a point $z$ such that $f_{0}(z)<z$,
so we can deduce that $(W_{n})_{n}$ is an infinite sequence and converges.
On the other hand, if $f_{0}$ is always above the line $y=x+\epsilon$
then it is possible to show that $(W_{n})_{n}$ will terminate at
some integer $H_{f}<\infty$. Since $f_{0}\geq f_{1}$ (cf. figures
(\ref{fig:3}) and (\ref{fig:Using}) ), studying $f_{1}$ can give
us a sufficient condition for $H_{f}$ to be finite: having $f_{1}-\mathbb{I}>\epsilon$
would be that sufficient condition.

Let us study $f_{1}$ on $D=\mathbb{R}^{+}\cap\mathcal{D}(f_{1})$.
Assuming again $a_{0}>b_{0}+1>1$ and $0<\alpha<1$, we have both
$f_{0,1}(1)>1$. Now we want to see when would $f_{1}-\mathbb{I}>\epsilon$
or not i.e. when we can not or can find $y\in D$ such that $f_{1}(y)<y$.
The derivative of $(f_{1}-\mathbb{I})$ is: 
\begin{equation}
\begin{array}{ccc}
(f_{1}-\mathbb{I})'(x) & = & \dfrac{(1-\alpha)b_{0}x+a_{0}-b_{0}x-(a_{0}-b_{0}x)^{2-\alpha}}{(a_{0}-b_{0}x)^{2-\alpha}}\\
 & = & \dfrac{a_{0}(1-\alpha)+\alpha(a_{0}-b_{0}x)-(a_{0}-b_{0}x)^{2-\alpha}}{(a_{0}-b_{0}x)^{2-\alpha}}
\end{array}\:.\label{eq:r-1}
\end{equation}
Its sign is determined by its numerator that we will call $P(x)$;
$P$ is defined on $\mathcal{D}(f_{1})$. 
\begin{equation}
P'(x)=b_{0}(-\alpha+(2-\alpha)(a_{0}-b_{0}x)^{1-\alpha})\label{eq:z-1}
\end{equation}

So $P$ is, at first, strictly increasing until it reaches its maximum
value at $x_{1}\equiv\frac{1}{b_{0}}\left[a_{0}-\left(\frac{\alpha}{2-\alpha}\right)^{\alpha}\right]$
then becomes strictly decreasing. In addition to that, because $2-\alpha>1$
we have $P(-\infty)<0$ and at the limit $x\rightarrow a_{0}/b_{0}$
we have $P(x)>0$. So we can now establish a variation table for $P$
and from there we deduce that there is a unique $x_{0}\in\mathcal{D}(f_{1})$
such that $P(x_{0})=0$ and $P(x)<0$ for $x<x_{0}$ while $P(x)>0$
for $x>x_{0}$. We can even specify a bit more the value of $x_{0}$:
using the fact that $a_{0}>1$ we get $P(0)=a_{0}-a_{0}^{2-\alpha}<0$,
furthermore $P(x_{1})>0$. The conclusion from these two facts is
$x_{0}\in(0,\:x_{1})$. (Remark: $x_{1}>\frac{a_{0}-1}{b_{0}}$ because
$\left(\frac{\alpha}{2-\alpha}\right)^{\alpha}<1$ and $a_{0}>b_{0}+1>1$.)
From the sign of $P$, we finally deduce the variation table of $f_{1}-\mathbb{I}$.
In conclusion, $f_{1}-\mathbb{I}$ reaches its minimum at $x_{0}\in(0,\:x_{1})$;
$x_{0}=\frac{a_{0}}{b_{0}}-\frac{X_{0}}{b_{0}}$ where $X_{0}$ is
the (unique) root in $\mathbb{R}^{+}$ of $a_{0}(1-\alpha)+\alpha X-X^{2-\alpha}$.
If $f_{1}(x_{0})-x_{0}=\epsilon$ with $\epsilon>0$ then $H_{f}$
is finite since it would imply $f_{1}-\mathbb{I}\geq\epsilon$ as
$x_{0}$ was the minimum of $f_{1}-\mathbb{I}$. So $f_{1}(x_{0})-x_{0}=\epsilon$
is the sufficient (and probably nearly necessary) condition for $H_{f}<\infty$
and it only requires a numerical determination of $X_{0}$ then a
calculation of $f_{1}(x_{0})=1+x_{0}/X_{0}$.

Up to now, we saw that $f_{0}$ being below the line $y=x$ for some
real in $\mathcal{D}(f_{0})$ implies the existence of a fixed point
and $H_{f}=\infty$. Then we have managed to find some sufficient
condition for it not to be the case. Now let us suppose we do have
a fixed point. Since $f_{0}(1)>1$ the fixed points should be located
after $1$. Considering how $f_{1}$ vary, it does not have more than
2 fixed points. We will assume the same for $f_{0}$; $(W_{n})_{n}$
will then converge toward the smallest fixed point of $f_{0}$ that
we will assume to be equal to the smallest fixed point of $f_{1}$
(both should be very close from each other).

%\bibliographystyle{./epj}
%\bibliography{../../Biblio_Tree_paper}

\end{document}